\documentclass[10pt, conference, letterpaper]{IEEEtran}
\IEEEoverridecommandlockouts
\usepackage{cite}
\usepackage{amsmath,amssymb,amsfonts}
\usepackage{graphicx}
\usepackage{subfigure}
\usepackage{textcomp}
\usepackage{xcolor}
\usepackage{algorithm}
\usepackage[noend]{algpseudocode}
\def\BibTeX{{\rm B\kern-.05em{\sc i\kern-.025em b}\kern-.08em
    T\kern-.1667em\lower.7ex\hbox{E}\kern-.125emX}}
\begin{document}

\title{Masking Host Identity on Internet: Encrypted TLS/SSL Handshake}

\author{\IEEEauthorblockN{Vinod S. Khandkar and Manjesh K. Hanawal}
	\IEEEauthorblockA{\textit{Industrial Engineering and Operations Research} \\
		\textit{Indian Institute of Technology Bombay, Mumbai, India}\\
		\{vinod.khandkar, mhanawal\}@iitb.ac.in}
}


\maketitle

\begin{abstract}
Network middle-boxes often classify the traffic flows on the Internet to perform traffic management or discriminate one traffic against the other. As the widespread adoption of HTTPS protocol has made it difficult to classify the traffic looking into the content field, one of the fields the middle-boxes look for is Server Name Indicator (SNI), which goes in plain text. SNI field contains information about the host and can, in turn, reveal the type of traffic. This paper presents a method to mask the server host identity by encrypting the SNI. We develop a simple method that completes the SSL/TLS connection establishment over two handshakes - the first handshake establishes a secure channel without sharing SNI information, and the second handshake shares the encrypted SNI. Our method makes it mandatory for fronting servers to always accept the handshake request without the SNI and respond with a valid SSL certificate. 

As there is no modification in already proven SSL/TLS encryption mechanism and processing of handshake messages, the new method enjoys all security benefits of existing secure channel establishment and needs no modification in existing routers/middle-boxes. Using customized client-server over the live Internet, we demonstrate the feasibility of our method.  Moreover, the impact analysis shows that the method adheres to almost all SSL/TLS related Internet standards requirements. 
\end{abstract}


\section{Introduction}
The advent of modern content delivery techniques has increased Internet traffic many folds over the last decade. As more users rely on Internet services, data privacy, and anonymity over the Internet have become crucial. This need prompted the rise of secured or encrypted data communication using HTTPS protocol \cite{rfc2818}. These days, most of the Internet services use the HTTPS protocol for providing data anonymity and security. 

While HTTPS protocol secures the channel content, it also makes the traffic stream unidentifiable by network middle-boxes. It has restricted Internet Service Providers' (ISPs) ability to efficiently classify the traffic streams that would help them perform legitimate tasks like traffic management or apply discriminatory policies like selective throttling and preferential treatment of specific traffic streams. Thus, it has prompted ISPs to employ various other techniques to classify traffic \cite{tr_cl}. The ``Server Name Identification" (SNI) based technique is one such commonly used technique. It uses the SNI field in the Transport Layer Security (TLS) handshake protocol message to identify the traffic flow. It is an optional parameter sent as a plain-text in the initial SSL/TLS handshake. The ``Domain name Service" (DNS) database associated with the network address is also a primary source of revealing traffic flow identity. However, with the adoption of ``DNS over TLS" \cite{rfc7858} techniques, the DNS information is encrypted over the Internet. That means it is not readily available to network middle-boxes for traffic flow identification. That leads to the SNI based traffic stream identification \cite{sni} as a widely used technique over the Internet.

\begin{figure}[htbp]
	\centerline{\includegraphics[width=2.5in]{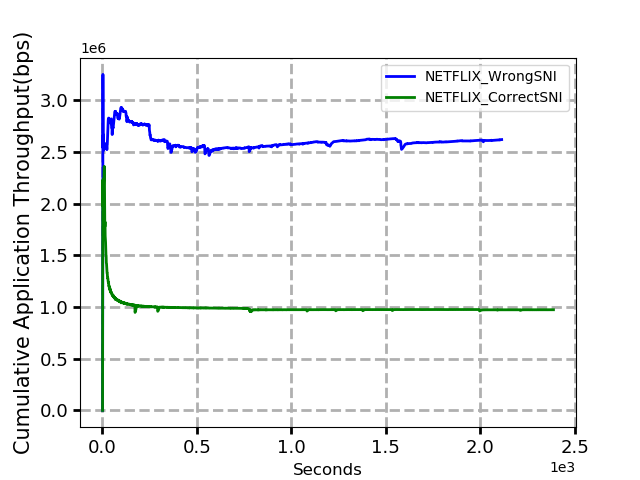}}
	\caption{SNI based traffic shaping illustrated using commercial traffic shaper}
	\label{fig:ts_raath}
\end{figure}

The network middle-boxes can manipulate throughput of traffic streams with the correctly identifiable class or server host-name. Often, the identification is made based on SNI information. Fig.~\ref{fig:ts_raath} illustrates SNI based traffic shaping using commercial traffic shaper. We use a commercial traffic shaper to throttle throughput of a video stream to $1$ Mbps. When the video stream access mechanism uses the correct SNI, the traffic shaper successfully limits the throughput to $1$ Mbps (green curve). When the same video access mechanism does not use any SNI information or uses non-comprehensible SNI, with everything else remaining the same, the stream gets throughput of $2.5$ Mbps based on bandwidth availability (blue curve). It illustrates that the commercial traffic shaper could correctly identify and throttle the video stream in the presence of correct SNI information. In the latter case, it could not classify the stream and hence did not throttled the throughput. It validates our claim that the SNI field is essential for the middle-boxes/traffic-shapers to correctly classify the traffic streams and apply any traffic management or throttling.

Even though ISP middle-boxes' service identification and classification aid efficient traffic management, it also gives far more control over the handling of packets related to specific services within the network. The ISP middle-boxes often use such control to provide higher performance to certain services while degrading certain services' performances. There are reports of preferential treatment cases in the past, e.g., Netflix v/s Comcast, that prompted various regulating authorities to advocate net-neutrality. The net-neutrality is related to `equal' treatment of all packets over the Internet. 

This paper proposes a method to send SNI in a non-plain-text format that prevents the middle-boxes from using it to identify the traffic class or server host-name. Thus, it paves the way for more comprehensive and non-malicious ways of managing traffic over the Internet. It will also help prevent malicious users from knowing the type of activity performed by users while using public networks through wire-tapping. It applies to all versions of TLS supporting SNI parameter. Our contributions can be summarized as follows:
\begin{itemize}
    \item We develop a method to mask SNI which conceals host's identity which in turn prevents classification of traffic streams on the Internet
    \item We validate our method works without requiring any changes in the existing routers/middle-boxes by establishing a secure connection between a server-client pair over the Internet.
\end{itemize}

The paper is organized as follows:
Section~\ref{sec:background} describes the background technologies associated with the proposed methodology. 
Section~\ref{sec:methodology} describes the design principles followed and the proposed method. Section~\ref{sec:phandshake} explains modification to the existing the TLS handshake to send SNI on a secure channel. Section~\ref{sec:sys_eval} discusses impacts of various design choices and validates the feasibility of our methods through its implementation on live Internet. The advantages and disadvantages of the proposed method are discussed in Section~\ref{sec:lim_advt} with a conclusion in Section~\ref{sec:conclusion}

\subsection{Related Work}
Three revisions of TLS protocol have been made since its introduction as an improved SSL protocol. The latest TLS version is 1.3. It was introduced in 2018. Apart from the major TLS/SSL releases, there have been many protocol and parameter related modifications proposed until now. The possibility of improving the basic RSA algorithm's speed is investigated using the EAMRSA algorithm in \cite{eamrsa_ssl}. \cite{isakmp_ssl} proposes the new key negotiation mechanism that is ISAKMP based. Integrating ISAKMP in SSL brings notably new authentication methods, identity protection, and fast algorithm negotiations. The attempt to encrypt/decrypt SSL/TLS handshake parameter is done in \cite{csk_ssl}. It proposes the Combined Symmetric Key (CSK) based SSL handshake that uses CSK technology to authenticate both communication sides and uses the symmetric key to encrypt and decrypt secret information, i.e., server authentication code. \cite{ssl_imp} introduces OTP on a mobile-based pre-SSL-handshake mechanism to tackle the issue of fake root certificate installed on the user device. The password generated in the pre-SSL-handshake phase is used to generate parameters for a regular SSL handshake. 

A standardization body acknowledges the need for encrypted SNI (E-SNI) with the related issues and requirements well documented in the internet-draft \cite{draft-ietf-tls-sni-encryption-09}. It lists unanticipated usage of SNI that includes ISP QoS profiling based on SNI. The various E-SNI requirements are categorized as attack prevention, fronting server related, and other security considerations. The E-SNI is still not available as an Internet standard, but the internet-draft in an experimental version is available \cite{draft-ietf-tls-esni-07}. It proposes to encrypt the entire ``clientHello" message. A slightly different version of it is already implemented and commercialized by ``Cloudflare" \cite{cf_esni} that encrypts only the SNI portion of the ``clientHello" message. It uses the Diffie-Hellman key exchange algorithm to generate a shared encryption key over an untrusted channel. This method's main drawback is that it requires a considerable amount of pre-handshake processing and requires the client and server to exchange the public portion of the generated symmetric key to encrypt and decrypt SNI. This requirement raises the security issues that need mitigation using persistent security credential modifications.  

The authors in \cite{tor} propose a method of domain fronting in which the fronting server establishes a secure connection with the client using its server name. Later this connection is extended to the intended content server by establishing another HTTPS tunnel within already established HTTPS tunnel. The major drawback of this method is the increased workload of double encryption. 

Our SNI encryption method is different from the above methods as it not only encrypts the SNI but performs the entire SSL handshake for appropriate SNI within a secured channel. It not only hides SNI from middle-boxes but also the plain-text server certificate. It is thus making the entire SSL/TLS handshake in question transparent to network middle-boxes.

\section{Background}
\label{sec:background}
The method proposed in this paper deals with SNI and SSL/TLS handshake, along with its operation in various client-server configurations. This section describes these aspects as defined in the standard and deployed in practice.

\subsection{SNI in TLS}
SNI is an optional field in the ``clientHello" message sent from the client. This message is part of the initial handshake of the Transport Layer Security (TLS) protocol \cite{rfc8446}. IETF standardized the TLS protocol by taking Secured Socket Layer (SSL) as a seed protocol. As of today, it is the most widely deployed protocol for establishing a secure channel. ``HTTP over TLS" or HTTPS standard has adopted the TLS protocol in which the TLS provides channel-oriented security to already defined plain-text HTTP protocol. 

\begin{figure}[htbp]
	\centerline{\includegraphics[scale=.51]{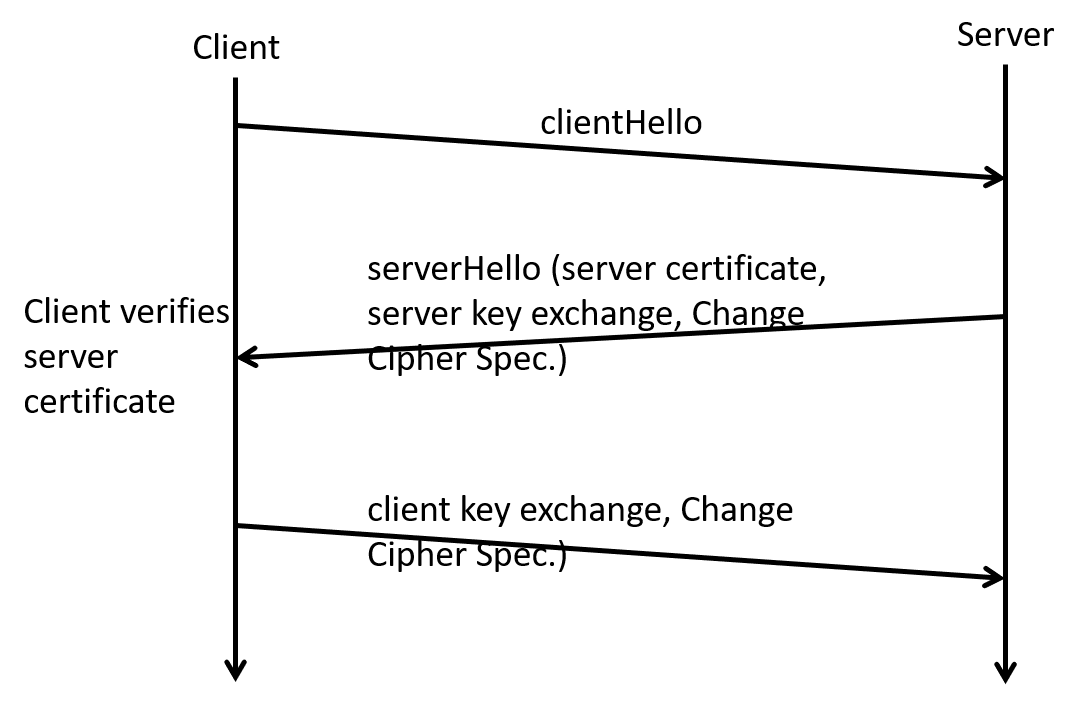}}
	\caption{TLS handshake}
	\label{fig:tls_hs}
\end{figure}

The handshake protocol of TLS consists of a ``clientHello" message from the user-client initiating establishment of a secure channel (see Fig.~\ref{fig:tls_hs}). The user-client generates TLS requests by using the information from the public URL of the content-server, e.g., https://www.youtube.com. This information contains the server name or SNI. Usually, the user-client includes this information in the ``clientHello" message as a ``server-name-indication" extension. \cite{rfc3546} has added it to the TLS handshake. This extension allows servers to host multiple `virtual' servers at a single underlying network address. Yet, the user-client can make distinct secure HTTPS connections towards individual virtual servers using their security certificate. If accepted, the server responds to the user-client request with the ``serverHello" message containing its certificate and other encryption-related information. The client is responsible for checking the authenticity of the certificate. The client also verifies the host-name in the received server certificate, thus thwarts any Main-in-the-Middle (MITM) attacks. 

Currently, network middle-boxes are well equipped to identify the Internet service using plain-text server name information present in TLS ``clientHello" message and sometimes in server certificate as part of the TLS ``serverHello" message.      

\subsection{TLS termination points}
Transport Layer Security or TLS is an end-to-end protocol. While it terminates in the user client on one side, on the other side, it can terminate in various network entities depending on the associated client-server configuration. We will discuss these configurations in detail in the remainder of this section.

\subsubsection{Directly accessible servers}
In this configuration, a physical network address maps to a single domain. There is no fixed intermediary device defined by the server for communicating with it. So the client directly establishes a socket connection with the server. After this, the client and server serve as a end-points for further communication, including the TLS handshake. The client can optionally include SNI in the ``clientHello" message. As the server hosts a single domain, it always responds with the same certificate in the ``serverHello" message. Thus, optionally, SNI in ``clientHello" message or plain-text server certificate reveals the server or related service identity in this scenario. 
     
\subsubsection{Servers with transparent proxy}
Sometimes, servers employ transparent proxies \cite{rfc1928, rfc3089, rfc1919} for using standard client-server procedures while taking advantage of proxy connection. Transparent proxies create two separate connections to support a single logical connection from the client to the server. It creates one connection from the client to proxy and second from proxy to server. As the name suggests, it transparently transfers packets from one connection to another. Thus it behaves like an IP gateway for application data. That means the end-to-end TLS connection is negotiated and established between client and server transparent to the proxy. Hence this configuration is similar to directly accessible server case in terms of SSL/TLS parameter visibility on the Internet. 

\subsubsection{Gateway server}
Gateway server or reverse proxy \cite{rfc7230} is an intermediary that acts as an interface towards the client for single or multiple domains or physical server machines. It is mainly used for `accelerator' caching and enables partitioning or load balancing of HTTP services across multiple machines. Thus single physical network address of the gateway server is associated with multiple domains or servers. It adheres to all standard HTTP connection requirements towards the client but can choose any desirable communication mechanism towards content servers. Thus the gateway server acts as another end-point for end-to-end TLS protocol handshake. All clients that wish to communicate with different domains or servers represented by the gateway server's network address, sends TLS ``clientHello" message to a single gateway. It is the gateway server's responsibility to select the correct server certificate. It performs this selection based on the SNI parameter in the ``clientHello" message. Thus, such server deployments critically need the SNI field, and its absence results in TLS handshake termination. It is currently the most widely used server configuration, and many types of dedicated gateways are standardized, e.g., Media Gateway \cite{rfc3525}.

Due to SNI's crucial role in such configurations, the services hosted are readily identifiable using the SSL/TLS handshake's SNI parameter. The domain name interfacing using the content delivery network's edge server falls in this category. The edge server acts as a Gateway server. In this case, the TLS handshake requires the domain owner to give their certificate's private key to the edge server. The security issue in handing over server certificate is solved using the keyless SSL mechanism \cite{keyless_ssl_patent} in which the certificate remains with the original domain, but ``clientHello" is still processed by the edge server with the help of the associated domain server. The Gateway server does not always hold the content server's security certificate but can retrieve it if required. Thus, the Gateway server's domains use the Gateway as a TLS end-point due to its capability to retrieve security certificates.

\subsubsection{Virtual Private Network (VPN)}
There are many types of VPN services defined by IETF \cite{rfc2764}. However, from the TLS protocol handshake point of view, they all can be considered one type. The reason is that Layer-2 layer security/encryption provided by VPN and TLS protocol runs above the transport layer protocol. All VPN packets are encapsulated and encrypted at the L2 layer. So the SNI field present in the TLS protocol message is inherently transmitted within encrypted L2 messages. This configuration does not put any requirement of the SNI encryption.


\subsection{TLS/SSL certificate}
TLS/SSL certificates \cite{ssl} are an electronic document that contains information about the authenticity of specific details associated with the issued entity, e.g., content-servers or services. Only trusted third party authorities can issue signed certificates. Such entities are called Certificate Authority (CA). The standard web accessing mechanisms such as web-browser always trust such signed certificates. It is also possible to generate the self-signed certificates using ``open-ssl" \cite{open_ssl} APIs. However, such certificates do not have validity across the Internet. 

SSL certificates contain information such as public key, domain name. These certificates can be specific to one or multiple domains, and the certificates stores this information. The other types of certificates are not specific to the domain but to the requesting entity.  Thus requesting entity can use such a certificates for any of its needs, such as server authenticity or digital signatures. During the SSL secure channel handshake, the server sends the certificate to the requesting entity. It is requesting entities' task to validate the certificate's contents to check the server's authenticity. 

\label{sec:into}

\section{Methodology}
\label{sec:methodology}
\begin{figure*}[!t]
	\centerline{\includegraphics[scale=.4]{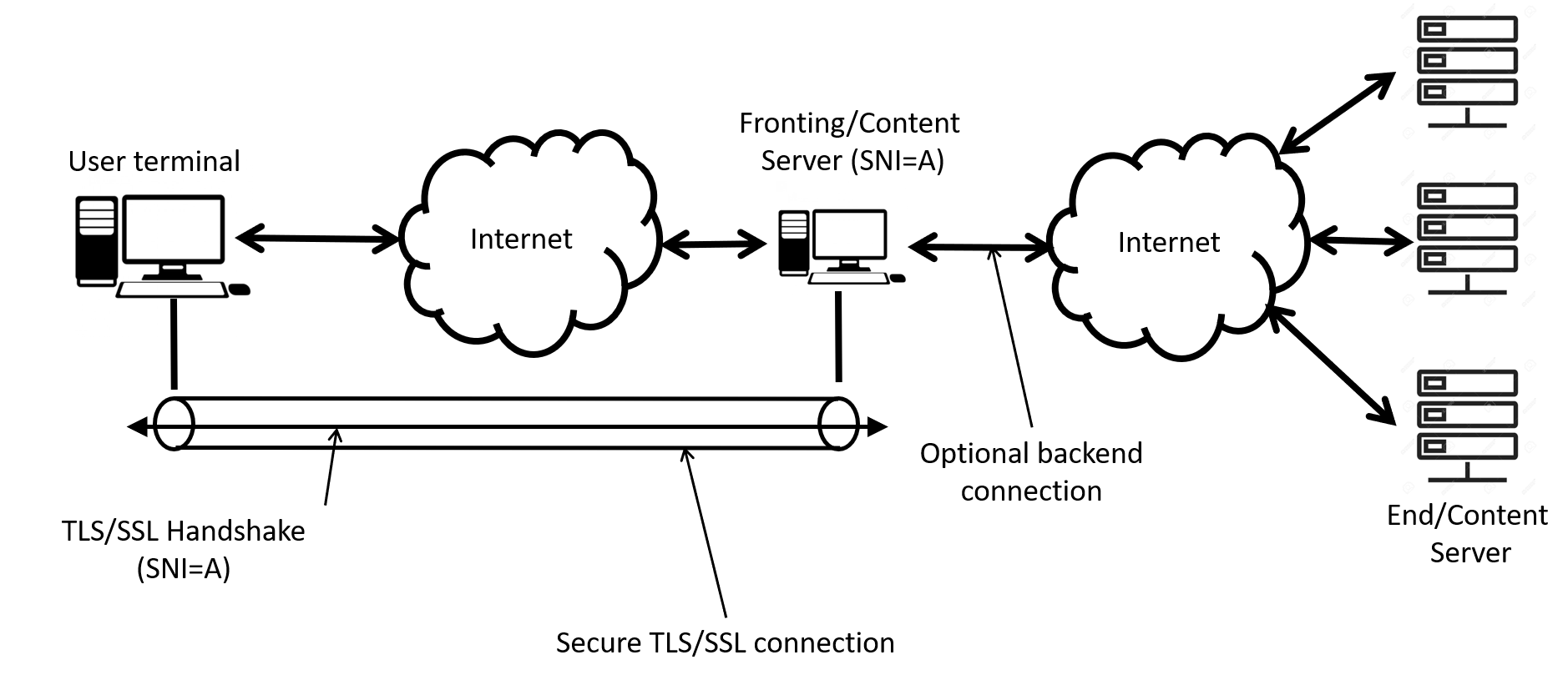}}
	\caption{Proposed methodology}
	\label{fig:method}
\end{figure*}
In this section we describe design principles behind the proposed solution and a methodology to encrypt SNI. It also discusses the flexibility and robustness of our method. 

\subsection{Solution principles}
We identify and follow a few solution principles while developing a solution. We discuss those next and then present our methodology that follows these principles.

\subsubsection{Same existing core algorithm} 
TLS/SSL secure channel mechanism has evolved into a robust and widely used mechanism over a long time. Any new method should not modify the message flow or the existing TLS/SSL mechanism's parameter contents to carry forward all its benefits. It should follow an add-on-feature design. Such a constraint will make our solution ready to be used with future versions of TLS/SSL.   

\subsubsection{Minimum encryption overhead}
The SNI encryption can add overhead in terms of processing required at both ends and the number of exchanged parameters. Any new solution should not add a considerable amount of such overheads, e.g., encryption parameter handshake or frequent peer-to-peer parameter exchanges. 

\subsubsection{Same number of secure tunnels}
The ``multiple tunnel-in-tunnel" approach leads to multiple encryptions of the same packets in overlaid layers. It is not desirable for delay-sensitive services. Any new solution should keep such overheads at a minimum. 

Even though our solution follows the principles mentioned above, it also fulfills many other requirements listed in \cite{draft-ietf-tls-sni-encryption-09}. A few are listed below,
\begin{enumerate}
    \item \textbf{Supporting multiple transport protocols:} Our solution modifies only TLS handshake. It means it can work with all transport layer protocols that work currently with TLS.
    \item \textbf{Hiding the Application Layer Protocol Negotiation:} Our solution encrypts all messages in SSL/TLS handshake sequence that uses intended SNI. It includes messages containing application layer negotiation information.
    \item \textbf{HTTP Co-Tenancy Fronting:} The mandatory requirement of default SSL server certificate makes our solution to support all server configurations, including Co-Tenancy Fronting. 
\end{enumerate}

\subsection{Proposed solution}
Our methodology is to perform the regular legacy TLS/SSL handshake over the already established secure channel. For that, the client and server first establish the secure channel without any SNI that identifies the traffic flow. The immediately following another handshake with the SNI of the intended server uses this secure channel. This second handshake modifies the server certificate and related ciphering parameters associated with the existing secure channel. Thus the SNI and server identifiable parameters that are part of legacy TLS/SSL handshake as plain-text are sent on the Internet as encrypted text, as shown in Fig~\ref{fig:method}. 

The TLS protocol supports the modification of the existing secure channel with the new server certificate and ciphering parameter. However, establishing an initial TLS/SSL secure channel without associating it with the content server, i.e., the intended SNI parameter, poses challenges. The main reason being the different types of client-server connection configurations possible. Next, we will discuss the issues associated with SNI values and different types of client-server configuration. 

The absence of SNI is not an issue for the servers with a one-to-one mapping between network address and domain. However, for servers employing co-tenancy, e.g., edge servers of the content delivery network, SNI information is crucial. The absence of SNI leads to termination of the TLS/SSL handshake due to the non-availability of the required certificate. So the SNI field cannot be empty without any support from the server-side for all client-server connection configuration.

The TLS/SSL is an end-to-end protocol. The secure channel end-points that also performs initial handshake are dependent on client-server configuration. While the client-side works as a fixed end-point, the server-side end-point can be a direct content server, gateway server, or some proxy. Moreover, the intended server certificate is also not guaranteed to be present at the current TLS/SSL negotiating end-point. ``keyless SSL" technology \cite{keyless_ssl_patent}  deliberately keeps it out of the fronting server. It needs the fronting server to fetch it depending on the SNI parameter information in ``clientHello" message. That means the TLS/SSL end-point on the server-side is independent of the server certificate location. There can be a series of devices between clients and servers capable of interacting with clients using TLS/SSL protocol. However, not all devices can retrieve the intended server certificate for participating in SSL/TLS handshake. It is impossible to modify existing TLS connection with the intended server certificate if SSL/TLS handshake is negotiated with nodes that are not capable of it. Such an attempt will result in establishing TLS/SSL secure channel with wrong end-points and wrong security credentials but exposing the SNI during the process.

As mentioned above, the proposed method faces challenges in setting and handling SNI and adding server-side support at an appropriate node on the server-side. Now, we will discuss how our method resolves these problems.

\subsubsection{SNI in the first handshake}
As described above, the SNI parameter's validity and the necessity for completing successful SSL/TLS handshake changes with the client-server configuration, e.g., while it can be optional for the direct server connection, it is crucial for Gateways to represent multiple domains. Our method removes this dependency by making it mandatory for the user-client to perform the first SSL/TLS handshake without SNI and server-side node to accept and successfully process such handshake requests with SNI absent. It guarantees successful acceptance of a handshake request even though it does not contain any SNI, and the fronting server hosts multiple domains. Many times DNS entry for any physical network address may not contain server name in the usable format defined by SNI extension requirements \cite{rfc6066}. It is also barred to use  IPv4/IPv6 addresses in the SNI extension directly. In such cases, not all fronting servers can provide their server name that can be directly usable in the SNI parameter of the TLS handshake. So the SNI parameter in the first TLS/SSL handshake request is made ``ABSENT" to handle all such scenarios.     

\subsubsection{Establishing secure channel for requests without SNI}
As per the proposed modification, the server-side must successfully establish the secure channel even if the corresponding request is received without SNI extension. To achieve this, server-side TLS/SSL end-point should have a valid SSL certificate that can be used in the absence of SNI in the request. It is the mandatory support required from the server-side.     
  
\subsubsection{Server-side TLS/SSL end-point}
As per the proposed method, the client first establishes the secure channel without the SNI parameter in handshake request and using the default SSL server certificate. This secure channel is used to perform another handshake that modifies the secure channel's security credentials by including the appropriate SNI parameter in the request. Such modification is possible only if both handshakes are performed with the same end-points. Thus the support for handling TLS/SSL handshake requests without SNI, i.e., valid default server certificate, needs to be installed on the device capable of retrieving the intended content server's security certificate. Note that the latter certificate is useful in the second handshake but not in first.

\section{Modified handshake}
\label{sec:phandshake}
Our method uses two TLS/SSL handshakes to send the SNI and negotiate security parameters securely. A client that wishes to establish a new HTTPS connection towards the given server first triggers the TLS/SSL handshake as usual. The same pre-defined SNI parameter is always present in the first TLS/SSL connection establishment irrespective of knowledge of underlying client-server connection configuration. The SNI parameter in this request is always absent (refer Algo.~\ref{client-side}). The existing client behavior is not modified for remaining messages in handshake procedures, e.g., ``serverHello" message processing or ``clientHelloDone" message.  

\begin{figure}[htbp]
	\centerline{\includegraphics[scale=.4]{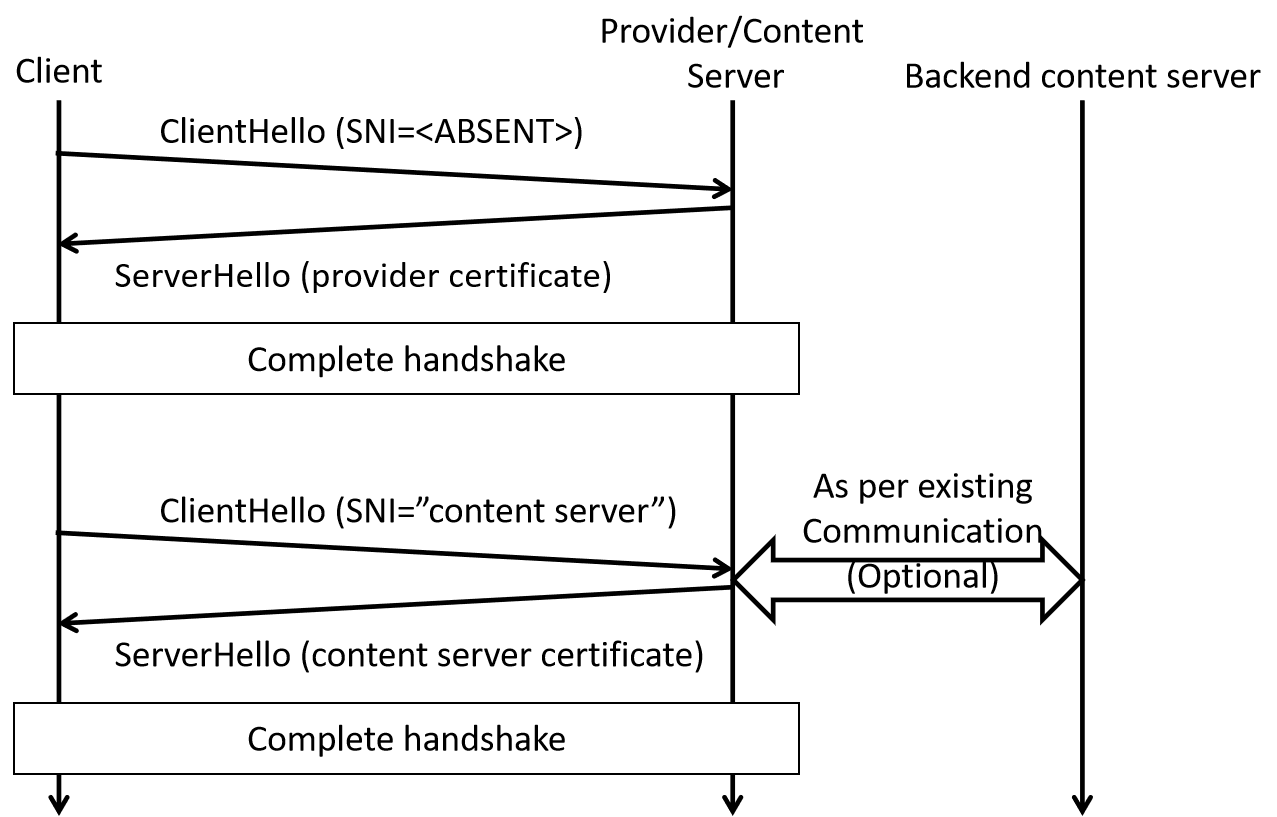}}
	\caption{Proposed TLS handshake}
	\label{fig:prop_tls_hs}
\end{figure}

\begin{algorithm}[!ht]
	\caption{Client-side SNI extension setting}\label{client-side}
	\begin{algorithmic}[1]
		\If{New secure channel establishment}
		\If {First handshake}
			\State Set SNI extension as absent
		\Else
			\State Set SNI extension as per existing setting
		\EndIf
		\Else
		\State Follow existing client behavior
		\EndIf
	\end{algorithmic}
\end{algorithm}

\begin{algorithm}[!ht]
	\caption{Server-side SNI extension processing}\label{server-side}
	\begin{algorithmic}[1]
		\If{New secure channel establishment}
		\If {First handshake}
			\If {SNI extension absent}
				\State Use default server SSL certificate
			\Else
				\State Follow existing server behavior
			\EndIf
		\Else
		\State Follow existing server behavior
		\EndIf
		\Else
		\State Follow existing server behavior
		\EndIf
	\end{algorithmic}
\end{algorithm}

On reception of the ``clientHello" message without the SNI parameter, the server accepts the incoming request. In such cases, it proceeds with the handshake using its default valid SSL certificate (refer Algo.~\ref{server-side}). This server behavior is different from the existing server response, as it always looks for valid SNI in case it hosts multiple domains. Currently, it has a provision to reject the incoming request message in such cases. The successful completion of the first handshake establishes a secure channel without any server identity associated with it. This phase is identical to all clients connecting to any server from the network middle-box's traffic flow identification point of view. Note that the server-side end-point performs the handshake process as usual if the SNI extension is present in the first handshake request. It is to ensure backward compatibility.   

Upon completing the first phase or handshake successfully, the client invokes a second handshake for the same connection. This handshake contains the regular protocol parameters and, more importantly, the intended SNI parameter in the request message. The already established secure channel ensures the data encryption of messages exchanged during the second handshake. The client and server complete the handshake as usual but over the secure channel. This second handshake modifies the already established secure connection to the credentials as per the server certificate associated with the server name in the SNI extension. 

Fig. \ref{fig:prop_tls_hs} shows the proposed TLS/SSL handshake mechanism. The location of the default server SSL certificate ensures that both handshakes use the same set of end-points. There is no change in the sequence of TLS/SSL handshake messages or its contents except for multiple handshakes and SNI restrictions from the existing one. Thus it provides all the security benefits of legacy secure channel establishment procedures.

The absence of the SNI in the first handshake procedure allows the edge server to use the same certificate for multiple domains or multiple server locations.

\section{System Evaluation}
\label{sec:sys_eval}
\begin{figure*}[btp]
	\centerline{\includegraphics[scale=.45]{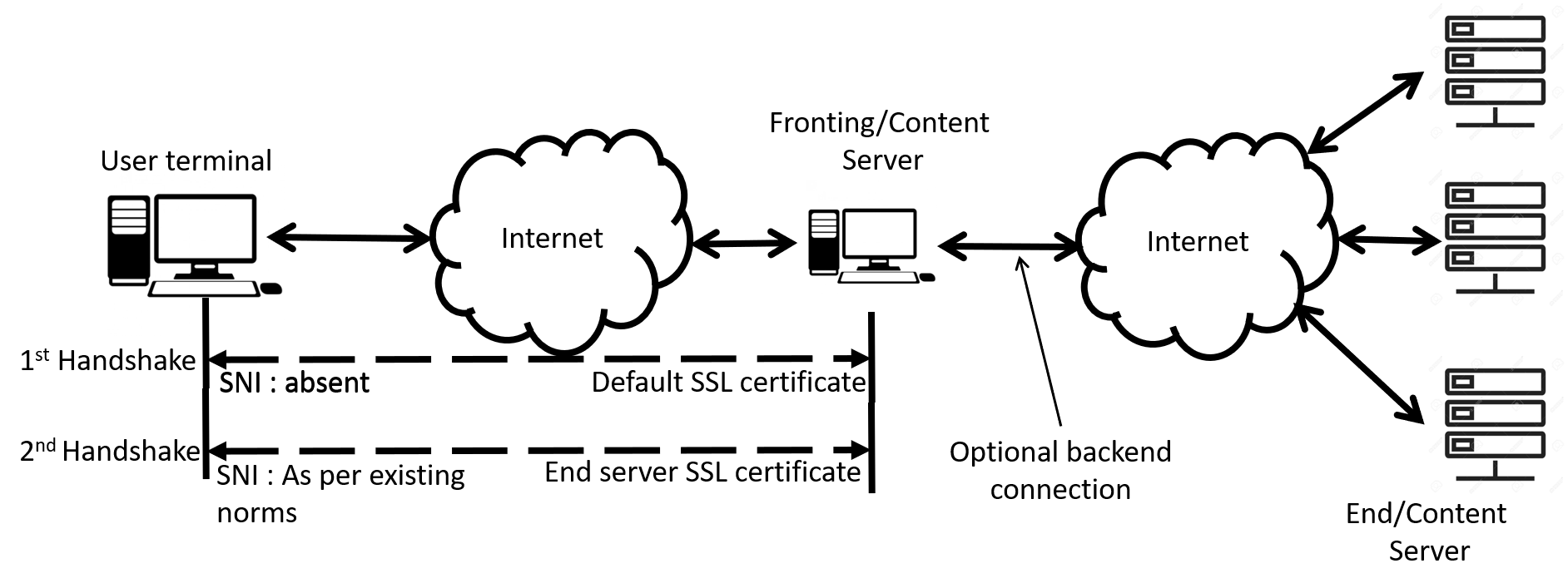}}
	\caption{Proposed modified end-to-end system}
	\label{fig:mod_sys}
\end{figure*}

Figure~\ref{fig:mod_sys} shows the end-to-end system with the modified TLS/SSL handshake protocol with associated end-points. The TLS/SSL handshake is a procedure between the user-client and fronting/content server. If different from the actual content server, this fronting server can be connected to the content server using a different secure channel. However, it is equipped with its valid SSL certificate and can retrieve content server related security credentials or its SSL certificate. These end-points perform two handshakes successively to establish a secure channel between client and server end-point with the content server's security credentials. 
 
Even though the SNI parameter is optional, the presence or absence impacts different server types differently. Moreover, the mandatory requirement of a server certificate at the fronting server and other modifications impacts system behavior defined by Internet standards. In this section, we will discuss such impacts generated due to newly added or modified special processing. We also validated the end-to-end feasibility of the proposed methodology over the Internet using customized client and server communicating over the live network. 

\subsection{End-to-end operation validation}
The end-to-end validation setup resembles the system in Fig.~\ref{fig:mod_sys}. The client and server communicate over the Internet. The client-server connection uses port number 443 for communication. For the demonstration purpose, the end-to-end system terminates at the fronting server. The fronting server hosts two SSL security certificates, both having different server names. The validation setup uses the self-signed SSL certificates. One certificate works as a default server certificate used in response to the ``clientHello" message with no SNI. 

During validation, the client first connects the server on port number 443.  It then performs the first TLS/SSL handshake without the SNI parameter in the ``clientHello" message. As a response to this handshake request, the server completes the handshake using its default certificate. The client validates it by verifying the server name in the received server certificate. After successfully validating the server certificate, the client performs the second TLS/SSL handshake with the pre-configured SNI parameter. The server receives the handshake messages as encrypted data as a result of transmission through a secure tunnel. At this point, the server recognizes the SNI in the incoming ``clientHello" message and responds with the associated second security certificate. The user-client validates the correctness of the server certificate by verifying the server name in it. 

\begin{figure*}[btp]
	\centerline{\includegraphics[scale=.28]{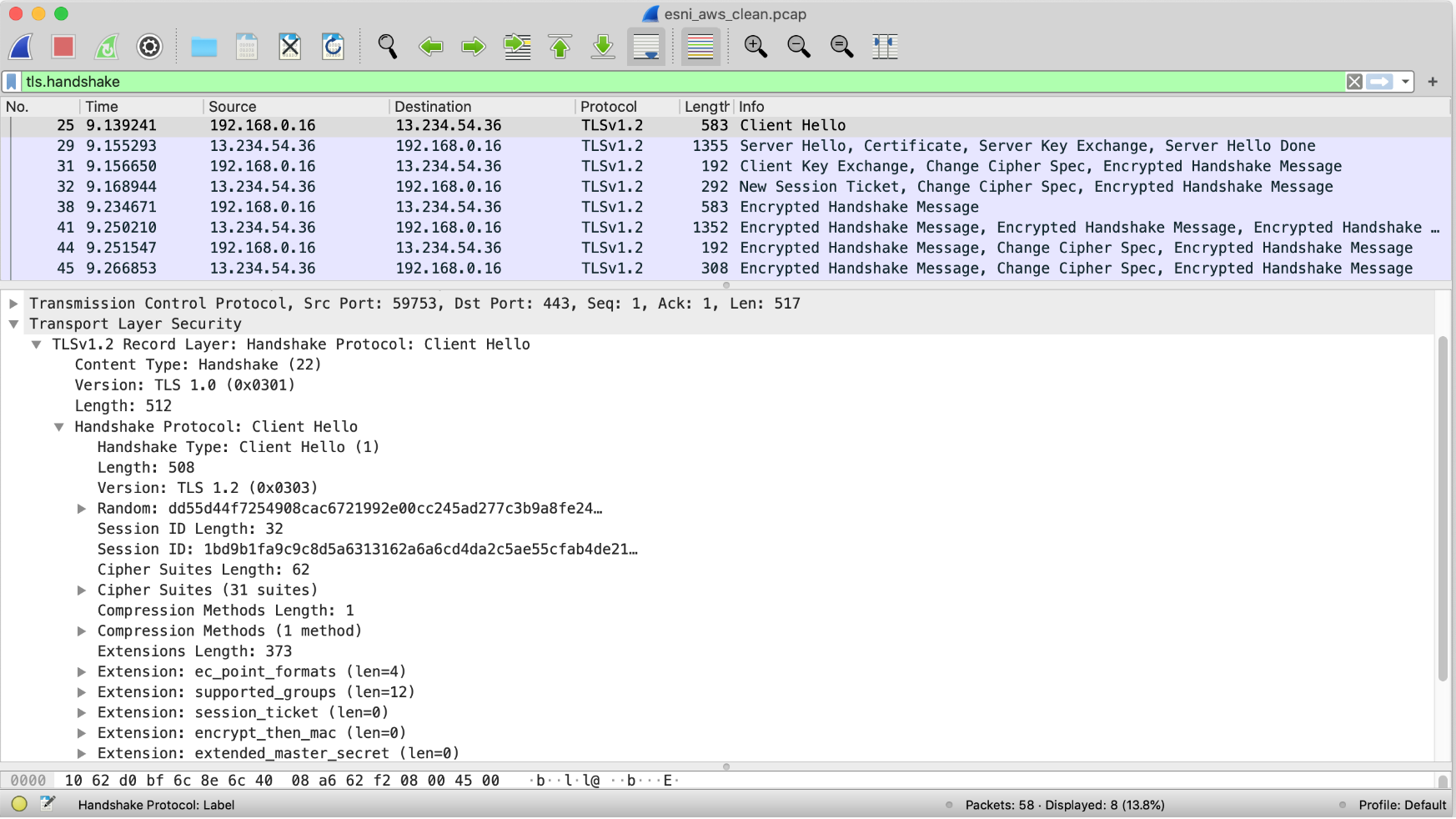}}
	\caption{Proposed modified TLS/SSL handshake demonstration}
	\label{fig:esni_demo}
\end{figure*}

Fig.~\ref{fig:esni_demo} shows the Wireshark logs for the end-to-end validation. The first four messages indexed by $25,29,31$ and $32$ corresponds to the first TLS/SSL handshake sequence, whereas the remaining messages indexed by $38,41,44$ and $45$ corresponds to TLS/SSL handshake with intended SNI value. As shown in the logs, all messages associated with the second handshake sequence are encrypted while traversing the Internet. 

\subsection{Processing overhead}
The proposed method modifies the original one-phase TLS/SSL handshake with a two-phase handshake. There is an additional processing overhead required to perform the second handshake. This overhead is quantified using the validation test of 100Mb data transfer.  The validation results show that this additional overhead amounts to 0.3--0.4$\%$ of total time duration. 

\subsection{Server Certificate}
Currently, the client-interfacing servers do not need to have a valid SSL certificate. The SNI parameter is an optional extension in the ``clientHello" message as it merely guides to select server certificate selection \cite{rfc5246}. The SNI field facilitates the secure connection to servers that host multiple ``virtual" servers at the underlying network address \cite{rfc6066}. Our method enforces the SNI field is absent even while negotiating TLS/SSL connection with such servers. The absence of SNI with the requirement of not failing the handshake due to SNI, demands such servers to have a default valid certificate. It is additional support required from the server-side end-point participating in the TLS/SSL handshake. Such certificates are not necessarily unique, but its deployment should adhere to other security requirements. Over re-use of the same certificate may attract security attacks on the server.  

\subsection{Client triggered ``second" handshake}
Usually, the user-client that has already established the secure connection uses another handshake to resume operation on the same secure channel \cite{rfc7525}. On resumption, the client should include the same SNI as used to establish the initial secure connection. Moreover, the server implementing the SNI extension "MUST NOT" accept the request to resume the session if the server$\_$name extension contains a different name \cite{rfc6066}. As per our method, the user-client always performs two handshakes - one with and another without the SNI extension. The server implementing the SNI extension should not consider the next immediate handshake as a resumption and accept the request even with different SNI.

\subsection{Application negotiated server name}
As per \cite{rfc6066}, the server name in the SNI extension, if present, should be the same as that of negotiated using application layer protocol before doing the SSL/TLS handshake. The servers should accept the first SSL/TLS handshake request without SNI that may differ from the application negotiated server name. The server should not fail a handshake due to such mismatch. 

\subsection{Common gateway interface}        
The ``Common gateway server" converts the user client's requests into respective ``Common Gateway Interface" (CGI) messages towards the intended server. It always copies the user client provided SNI value into the CGI interface without any processing logic. Moreover, it does not actively participate in the TLS/SSL handshake towards the user-client side in the form of feedback. So there is no special requirement from the "common gateway" interface processing. 

\subsection{Security}
The new handshake mechanism neither change the handshake end-points nor results in modification of other TLS/SSL handshake messages for exchanging handshake related information. So the security goals of TLS/SSL handshake are not compromised with the new handshake mechanism.

\section{Limitations and Advantages}
\label{sec:lim_advt}
In this section, we discuss the limitations and advantages of our method.
\subsection{Operational Limitations}
There is no change in the handshake mechanism from the message exchange sequence within a single handshake point of view. Nevertheless, specific changes in the handshake parameter setting limit the operational viability of the solution.
\subsubsection{Mandatory server certificate}
The client always sends the first TLS/SSL negotiation request without SNI. So the servers hosting multiple "virtual" domains with the same network address can not get the required SSL certificate from the hosted domain. Instead, it should have a valid certificate to respond to such requests. Even though multiple network addresses can use the same certificate, the additional certificate requirement and associated processing needed from the client-interfacing server limit the actual field deployment.

\subsubsection{SNI on back-end interface}
The SNI is encrypted over the interface from the user-client to the fronting server in our method. However, it can still be plain-text on the interface between the fronting server and the content server. 

\subsection{Advantages} 
The design principles followed to design the solution has resulted in many advantages.

\subsubsection{Backward compatible ready}
The solution is not entirely backward compatible, but it is ready for it. The server that wants to run in the legacy handshake mode only needs to suppress first handshake requests by generating appropriate return cause for user-client. On reception of such return cause, the user-client can terminate the current handshake and fall back to the legacy mechanism. We recommend the Internet standard to implement such a return cause for our method to be backward compatible.

\subsubsection{Same core TLS/SSL mechanism}
The core delicate TLS/SSL mechanism of certificate exchange, key generation, algorithm negotiation, and encryption is not changed. Thus there is no impact on the existing channel security.

\subsubsection{No proprietary encryption or message handling}
Our method uses encryption and handshake message processing mechanisms from the existing Internet standard. It does not contain any proprietary processing in terms of encryption or message processing. It makes the new mechanism to enjoy all benefits of existing TLS/SSL secure channel establishment. 

\section{conclusion}
\label{sec:conclusion}
The traffic stream identification by network middle-boxes is a significant breach/flaw in data anonymity over the Internet. It is acknowledged by the Internet standards committee as well. There are many methods by which internet middle-boxes can identify traffic streams. SNI based traffic identification is one such technique.  

This paper presents the method to mask the server host identity by sending the SNI as part of encrypted data over the Internet. The method consists of an SSL/TLS handshake in two phases --One without SNI and another with the intended SNI. It makes mandatory for fronting servers to always accept the handshake request without the SNI and respond with a valid SSL certificate. Establishing the secure channel with the first handshake ensures the encrypted transmission of all handshake messages during the second handshake. The second handshake then modifies the security credential of an already existing secure channel as per the intended content server.

As there is no modification in already proven SSL/TLS encryption mechanism and processing of handshake messages, the new method enjoys all security benefits of existing secure channel establishment. The methodology demonstration using the customized client-server over the live Internet confirms its feasibility. Moreover, the impact analysis shows that the method adheres to almost all SSL/TLS related Internet standards requirements.  

\bibliographystyle{IEEEtran}
\bibliography{IEEEabrv,vinod_bibfile}

\end{document}